\documentclass[9pt,twocolumn,twoside]{opticajnl}
\journal{opticajournal} 

\setboolean{shortarticle}{false}

\usepackage{lineno}

\title{On-chip measurement of the modal Stokes-Gell-Mann parameters for partially coherent three-mode light}
\author[1,$\dagger$]{Amin Hashemi}
\author[1,$\dagger$]{Abbas Shiri}
\author[1]{Bahaa E. A. Saleh}
\author[1]{Andrea Blanco-Redondo}
\author[1,*]{Ayman F. Abouraddy}

\affil[$\dagger$]{These authors contributed equally to this work}
\affil[1]{CREOL, The College of Optics \& Photonics, University of Central Florida, Orlando, Florida 32816, USA}

\affil[*]{raddy@creol.ucf.edu}

\begin{abstract}
The Stokes parameters are three real parameters that completely characterize partially coherent optical fields spanned by two modes -- whether a pair of polarization or spatial modes -- and their use is thus ubiquitous in optics. Because the Stokes parameters are defined through an expansion of the $2\times2$ coherence matrix in terms of the Pauli matrices, they cannot be applied to optical fields comprising three modes, which are described by a $3\times3$ coherence matrix. Examples of such fields include the polarization of non-paraxial fields (spanned by three orthogonal polarization modes), and fields comprising three spatial or temporal modes. It has long been theorized that the $3\times3$ Gell-Mann matrices -- developed in high-energy particle physics -- can serve as a basis for $3\times3$ optical coherence matrices, with 8~expansion coefficients known as the Stokes-Gell-Mann (SGM) parameters, but the measurement procedure is daunting, and the SGM parameters have not been measured directly to date in optics. Here we present the first measurements of the SGM parameters for partially coherent three-mode light in a photonic integrated platform comprising a hexagonal mesh of Mach-Zehnder interferometers. Measuring the SGM parameters on chip, from which we reconstruct the $3\times3$ coherence matrix facilitates exploring the full space of iso-entropy fields that can be inter-converted into each other unitarily, and those that share the same value of entropy and yet cannot be inter-converted unitarily. These results pave the way to utilizing multimode partially coherent light in applications involving optical communications, sensing, and information processing.
\end{abstract}

\setboolean{displaycopyright}{false} 

\begin{document}

\maketitle

\section{Introduction}

The manipulation of partially coherent light in integrated photonic platforms is emerging as a new frontier for optical information processing. This is motivated in particular by applications in which partially coherent light outperforms coherent light. Examples of this `coherence advantage' include utilizing partially coherent light in dense optical communications \cite{Nardi22OL}, scattering-free communications over a rapidly varying, strongly scattering optical channel without recourse to adaptive optics \cite{Harling25APLP}, improving the parallelization of on-chip optical computing \cite{Dong24Nature}, optical  cryptography \cite{Peng21P,Liu25LPR}, and spectroscopy \cite{Miller25Optica}. These applications typically make use of optical fields spanned by a finite number~$N$ of modes, but with random variables as the modal coefficients, whereupon the field characteristics can be tabulated in an $N\times N$ coherence matrix \cite{Gamo64PO} rather than a continuous coherence function \cite{Mandel65RMP,Wolf07Book}. We refer to such field configurations as `structured coherence'. Underpinning the applications of structured coherence in optical communications and information processing is the fact that an $N\times N$ coherence matrix requires $\mathcal{O}(N^{2})$ real parameters for its unique identification, in contrast to $\mathcal{O}(N)$ real parameters for its coherent counterpart \cite{Gamo64PO,Waller12NP}. Whereas this unique feature provides the potential for denser information encoding, it concurrently makes the measurement process challenging \cite{Waller12NP}. Indeed, the reconstruction of an $N\times N$ coherence matrix $\mathbf{G}$ is currently emerging as a key challenge for exploiting structured coherence.

Because any reconstruction strategy of the coherence matrix will incorporate a potentially large array of cascaded interferometers, \textit{on-chip} implementations that provide the requisite scalability and stability for such a task \cite{Bogaerts20Nature} are currently being keenly pursued \cite{roques2024Light,Hashemi26arxiv,Hashemi26arxiv4Modes,Mor26arxiv}. One approach tomographically reconstructs $\mathbf{G}$ by acquiring modal Stokes parameters (SPs) as an intermediary \cite{Abouraddy14OL,Kagalwala15SR,Harling24PRA,Harling24PRA2,Harling25APLP,Hashemi26arxiv,Hashemi26arxiv4Modes}. Because conventional SPs correspond to the expansion coefficients of a $2\times2$ coherence matrix in terms of Pauli matrices \cite{Abouraddy14OL,Kagalwala15SR}, Stokes tomography can be extended to larger coherence matrices of dimensions $N\times N$ by identifying appropriate modal sets that span the space of Hermitian $N\times N$ matrices \cite{Hashemi26arxiv4Modes}. Another approach for reconstructing $\mathbf{G}$ relies on searching through a two-dimensional parameter space of unitaries implemented for each pair of modes \cite{roques2024Light,Mor26arxiv}. Although interesting, such an approach is significantly slower than Stokes tomography and may thus be unsuitable for optical communications and other high-speed applications.

To date, on-chip Stokes tomography has been realized with two-mode and four-mode partially coherent light to facilitate the reconstruction of the associated $2\times2$ \cite{Hashemi26arxiv} and $4\times4$ \cite{Hashemi26arxiv4Modes} coherence matrices, respectively. A challenge nevertheless remains for the reconstruction of \textit{odd}-dimensional coherence matrices. This problem also emerges for optical polarization in the non-paraxial regime where three orthogonal polarization modes are relevant \cite{Ramachandran80Pramana,Carozzi00PRE,Setala02PRE,Dennis04JOA,Ellis05OC,Ellis05PRL,Abouraddy06PRL,Refregier06JOSAA,Sheppard11JOSAA,Sheppard14PRA,Sheppard16JOSAA,Gil17PRA,ALonso23AOP}. Whereas the conventional SPs help reconstruct partially polarized \textit{paraxial} light spanned by two orthogonal polarization modes \cite{Brosseau98Book,Brosseau06PO}, a larger set of SPs is needed for its \textit{non-paraxial} counterpart. It has been suggested theoretically that the Gell-Mann matrices (developed in particle physics \cite{GellMann62PR}) can serve as a basis to define `Stokes-Gell-Mann' (SGM) parameters that fully characterize non-paraxial partially coherent light. The 8~Gell-Mann matrices, which are the generators of the SU(3) group, can be viewed as an extension to three-mode fields of the Pauli matrices, which are generators of the SU(2) group. However, rather than the 3~conventional SPs associated with the $2\times2$ Pauli matrices, the 8~SGM parameters are associated with the $3\times3$ Gell-Mann matrices. Despite considerable theoretical work done in optics on the SGM parameters for polarization modes, they have yet to be measured directly. Exceptions in the context of non-paraxial polarization include scattering the near-field light with three nano-scale dipoles to obtain linearly independent measurements to reconstruct the $3\times3$ polarization coherence matrix (strictly speaking, without recording the SGM parameters) \cite{Ellis05PRL}, and more recently computationally extracting the SGM parameters from the \textit{spatially} structured field at the back focal plane of a high-numerical-aperture lens \cite{Herrera25arxiv}.

\begin{figure*}[t!]
\centering
\includegraphics[width=18cm]{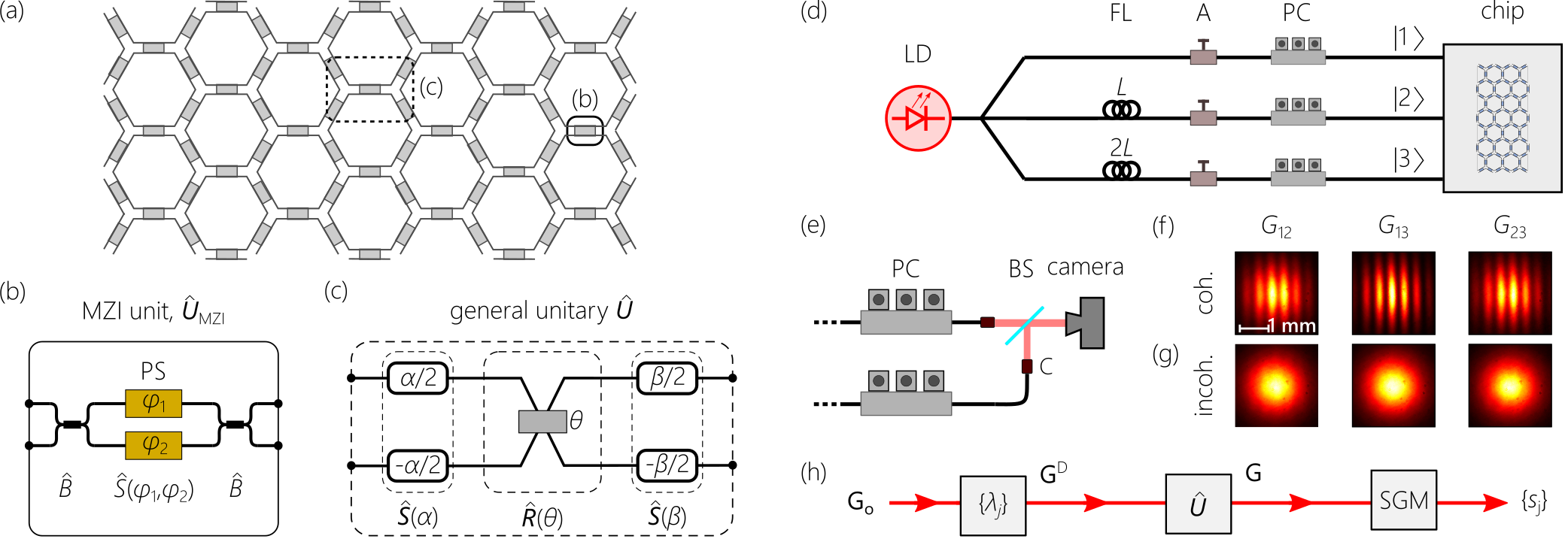}
\caption{(a) Layout of the hexagonal mesh of MZIs in the photonic integrated circuit used in our experiments. Each gray rectangle is an MZI, and the connecting lines are on-chip single-mode waveguides. The solid rectangle highlights a single MZI whose structure is elucidated in (b), and a dashed rectangle highlights a portion of the circuit that corresponds to a general $2\times2$ unitary (e.g., the restricted unitary in Eq.~\ref{eq:RestrictedUnitary}) whose structure is elucidated in (c). (b) The structure of an individual MZI, which is formed of two phase shifters $\varphi_{1}$ and $\varphi_{2}$ on a pair of waveguides represented by the operator $\hat{S}(\varphi_{1},\varphi_{2})$, preceded and followed by symmetric couplers $\hat{B}$. (c) A general $2\times2$ unitary is formed of an MZI preceded and followed by phase shifters $\hat{S}(\alpha)$ and $\hat{S}(\beta)$. (d) Schematic of the optical setup. LD: Laser diode; FL: fiber loop; A: attenuator; PC: polarization controller. All lines are SMFs. The power from the LD is divided equally between the three SMF paths, which are coupled to the photonic chip, corresponding to the modes $|1\rangle$, $|2\rangle$, and $|3\rangle$. (e) Out-coupling of the fields from a pair of SMFs to record their interference visibility. C: Fiber coupler; BS: balanced beam splitter. (f,g) Images recorded by the camera in (e) after superposing a pair of fields from SMFs. (f) In absence of the fiber loops, the fields are mutually coherent and high-visibility fringes are observed for every pair of modes. (g) In presence of the fiber loops, the fringes are eliminated, and all modal pairs are mutually incoherent. The titles $G_{12},G_{13}$, and $G_{23}$ correspond to the off-diagonal elements of the $3\times3$ coherence matrix $\mathbf{G}$, with indices indicating the pair of modes used. (h) Sketch of the sequence of tasks that are accomplished on chip.}
\label{fig:OnChipMesh}
\end{figure*}

Here we realize an on-chip measurement scheme to obtain the SGM parameters for three-mode partially coherent light. The modes we investigate here correspond to the fields in single-mode on-chip waveguides and in single-mode fibers off-chip. Once the SGM parameters are determined, they enable reconstructing the associated $3\times3$ coherence matrix $\mathbf{G}$. We carry out our measurements in an integrated photonic platform comprising a hexagonal mesh of Mach-Zehnder interferometers (MZIs), followed by intensity measurements to extract the SGM parameters. To the best of our knowledge, this is the first time that the modal SGM parameters have been measured in optics whether in free space or on chip. We tune the field entropy from $S=0$ corresponding to coherent fields free of random fluctuations to the maximum entropy of $S=\log_{2}3$~bits for a completely incoherent three-mode field. We explore the space of iso-entropy states of three-mode light, which denotes the families of fields that share the same value of entropy $S$. Contrary to their two-mode counterparts \cite{Brosseau06PO}, not all iso-entropy three-mode fields can be inter-converted into each other via unitary transformations. We produce here both categories of iso-entropy three-mode fields: those that \textit{can} be inter-converted into each other unitarily, and those that \textit{cannot}. These results are expected to be useful in the realization of optical communications schemes that exploit partially coherent light \cite{Nardi22OL,Harling25APLP}, in optical sensing and spectroscopy \cite{Miller25Optica}, and in Stokes polarimetry with non-paraxial fields \cite{ALonso23AOP}.

\section{Measurement configuration}

We consider here three optical modes denoted by $|1\rangle$, $|2\rangle$, and $|3\rangle$ in the Dirac notation. A partially coherent field supported by these three modes is described by a $3\times3$ coherence matrix $\mathbf{G}$:
\begin{equation}\label{eq:3x3G}
\mathbf{G}=\left(\begin{array}{ccc}G_{11}&G_{12}&G_{13}\\G_{21}&G_{22}&G_{23}\\G_{31}&G_{32}&G_{33}\end{array}\right),
\end{equation}
where $G_{jk}=\langle E_{j}E_{k}^{*}\rangle$, $j,k=1,2,3$, $E_{j}$  is the field component associated with the mode $|j\rangle$, and $\langle\cdot\rangle$ corresponds to an ensemble average. The coherence matrix is Hermitian, $\mathbf{G}^{\dagger}=\mathbf{G}$, so that $G_{jj}$ is real, $G_{jk}=G_{kj}^{*}$, and its eigenvalues $\{\lambda_{1},\lambda_{2},\lambda_{3}\}$are real. We normalize $\mathbf{G}$ to unity trace, $\mathrm{Tr}\{\mathbf{G}\}=\sum_{j=1}^{3}G_{jj}=\sum_{j=1}^{3}\lambda_{j}$. Moreover, $\mathbf{G}$ is positive semi-definite, $0\leq\lambda_{1},\lambda_{2},\lambda_{3}\leq1$.

\subsection{Integrated photonics platform}

In our experiments we make use of a photonic integrated circuit (iPronics Smartlight Processor) to process partially coherent light. The chip comprises a hexagonal mesh of MZIs connected by an array of single-mode waveguides [Fig.~\ref{fig:OnChipMesh}(a)]. Each waveguide corresponds to one mode, and each MZI operates on a pair of such modes. An MZI is formed of two phase shifters imparting phases $\varphi_{1}$ and $\varphi_{2}$ to the two modes, corresponding to the operator $\hat{S}(\varphi_{1},\varphi_{2})=\left(\begin{array}{cc}e^{i\varphi_{1}}&0\\0&e^{i\varphi_{2}}\end{array}\right)$, preceded and followed by symmetric couplers corresponding to the operator $\hat{B}=\tfrac{1}{\sqrt{2}}\left(\begin{array}{cc}1&i\\i&1\end{array}\right)$. The resulting MZI described by the unitary transformation (`unitary' henceforth for brevity):
\begin{equation}\label{eq:MZI}
\hat{U}_{\mathrm{MZI}}=ie^{i\varphi}\left(\begin{array}{cc}\sin\tfrac{\delta}{2}&\cos\tfrac{\delta}{2}\\\cos\tfrac{\delta}{2}&-\sin\tfrac{\delta}{2}\end{array}\right),
\end{equation}
where $\delta=\varphi_{1}-\varphi_{2}$ and $\varphi=\tfrac{1}{2}(\varphi_{1}+\varphi_{2})$ [Fig.~\ref{fig:OnChipMesh}(b)].

Our scheme for measuring the SGM parameters makes use of unitaries operating on pairs of modes. The most general $2\times2$ unitary takes the form:
\begin{equation}
\hat{U}=ie^{\varphi}\left(\begin{array}{cc}e^{i\chi_{1}}\sin\tfrac{\delta}{2}&e^{-i\chi_{2}}\cos\tfrac{\delta}{2}\\e^{i\chi_{2}}\cos\tfrac{\delta}{2}&e^{-i\chi_{1}}\sin\tfrac{\delta}{2}\end{array}\right),
\end{equation}
which can be produced from $\hat{U}_{\mathrm{MZI}}$ by introducing phase differences $\alpha$ and $\beta$ between the modes before and after the MZI, respectively, where $\alpha=\chi_{1}+\chi_{2}$ and $\beta=\chi_{1}-\chi_{2}$ [Fig.~\ref{fig:OnChipMesh}(c)]. Setting $\beta=-\alpha$ yields a restricted unitary of the form:
\begin{equation}\label{eq:RestrictedUnitary}
\hat{U}=ie^{\varphi}\left(\begin{array}{cc}\sin\tfrac{\delta}{2}&e^{-i\alpha}\cos\tfrac{\delta}{2}\\e^{i\alpha}\cos\tfrac{\delta}{2}&\sin\tfrac{\delta}{2}\end{array}\right).
\end{equation}
For Hermitian matrices (such as the coherence matrix $\mathbf{G}$), the restricted unitary in Eq.~\ref{eq:RestrictedUnitary} suffices for most needed operations.

\subsection{Optical setup}

The optical setup used in our experiment is depicted in Fig.~\ref{fig:OnChipMesh}(d). The optical source is a fiber-coupled laser diode at a wavelength $\approx1.5$~$\mu$m with a finite linewidth that is split equally into three single-mode fibers (SMFs) that define the three modes in our experiments, $|1\rangle$, $|2\rangle$, and $|3\rangle$. Fiber loops of lengths exceeding the coherence length of the laser diode are inserted into the paths to render them mutually incoherent, and an attenuator and a polarization controller are inserted into each path to optimally couple the fiber mode to the chip.

We confirm that the modes in the SMFs are mutually uncorrelated before coupling to the chip by first out-coupling the fields to free space via fiber collimators, overlapping pairs of fields at a balanced beam splitter, and observing the intensity profile at a camera [Fig.~\ref{fig:OnChipMesh}(e)]. In absence of the fiber loops, we observe high-visibility fringes for any of the three pairs of modes [Fig.~\ref{fig:OnChipMesh}(f)]. In presence of the fiber loops, the fringes disappear [Fig.~\ref{fig:OnChipMesh}(g)], indicating that the modes are mutually incoherent. Because the power is equal in all three modes and all modal pairs are mutually incoherent, the $3\times3$ coherence matrix for this source is $\mathbf{G}_{\mathrm{o}}=\tfrac{1}{3}\hat{\mathbb{I}}_{3}$, where $\hat{\mathbb{I}}_{3}$ is the $3\times3$ identity matrix. 

\subsection{On-chip tasks}

Once coupled to the chip, a sequence of on-chip tasks are implemented, which are sketched in Fig.~\ref{fig:OnChipMesh}(h). First, the three eigenvalues $\{\lambda_{j}\}$, $j=1,2,3$, of the coherence matrix are tuned independently via non-unitary transformations. Each mode is directed to a single input port of an MZI with zero field provided to the other input. The output is extracted from only one port with the other output port discarded. Consequently, the power in the mode is reduced by a factor of $\sin^{2}\tfrac{\delta}{2}$ (Eq.~\ref{eq:MZI}). We use three such MZIs in the path of modes $|1\rangle$, $|2\rangle$, and $|3\rangle$ to introduce power reduction factors of $\eta_{1}=\sin^{2}\tfrac{\delta_{1}}{2}$, $\eta_{2}=\sin^{2}\tfrac{\delta_{2}}{2}$, and $\eta_{3}=\sin^{2}\tfrac{\delta_{3}}{2}$, respectively, thus yielding a diagonal coherence matrix $\mathbf{G}^{\mathrm{D}}$ of the form:
\begin{equation}
\mathbf{G}^{\mathrm{D}}=\left(\begin{array}{ccc}\lambda_{1}&0&0\\0&\lambda_{2}&0\\0&0&\lambda_{3}\end{array}\right)=\mathrm{diag}\{\lambda_{1},\lambda_{2},\lambda_{3}\},
\end{equation}
where $\lambda_{j}=\tfrac{1}{\eta_{1}+\eta_{2}+\eta_{3}}\eta_{j}$, $j=1,2,3$, $\lambda_{1}+\lambda_{2}+\lambda_{3}=1$, and $\mathrm{diag}\{\cdot\}$ is shorthand for a diagonal matrix with the entries representing its diagonal elements. Defining the field entropy $S=-\mathrm{Tr}\{\mathbf{G}\log_{2}\mathbf{G}\}=-\sum_{j=1}^{3}\lambda_{j}\log_{2}\lambda_{j}$ \cite{Gamo64PO}, this step allows us to tune the entropy of the field from $S=0$ when $\lambda_{1}=1$ and $\lambda_{2}=\lambda_{3}=0$ ($\delta_{1}=\pi$, $\delta_{2}=\delta_{2}=0$), which corresponds to a coherent field free of random fluctuations, all the way to a maximum entropy of $S=\log_{2}3$~bits when $\lambda_{1}=\lambda_{2}=\lambda_{3}=\tfrac{1}{3}$ ($\delta_{1}=\delta_{2}=\delta_{3}=0.392\pi$), corresponding to a maximally incoherent field (which is the input field $\mathbf{G}_{\mathrm{o}}$ coupled to the chip in our case).

Up to this point the coherence matrix is diagonal. The \textit{second} on-chip task is to mold the structure of $\mathbf{G}^{\mathrm{D}}$ via an arbitrary $3\times3$ unitary $\hat{U}$, resulting in a transformed coherence matrix $\mathbf{G}=\hat{U}\mathbf{G}^{\mathrm{D}}\hat{U}^{\dagger}$ in which prescribed off-diagonal elements are realized. The \textit{third} task is to measure the modal SGM parameters, which is accomplished by implementing a sequence of appropriate unitaries on the field followed in each of step of the sequence by modal-power measurements via on-chip detectors, from which we extract the SGM parameters. Using the measured SGM parameters we then reconstruct the synthesized coherence matrices to confirm their structure.

\section{Unitaries on three-mode light}

\begin{figure}[t!]
\centering
\includegraphics[width=8.4cm]{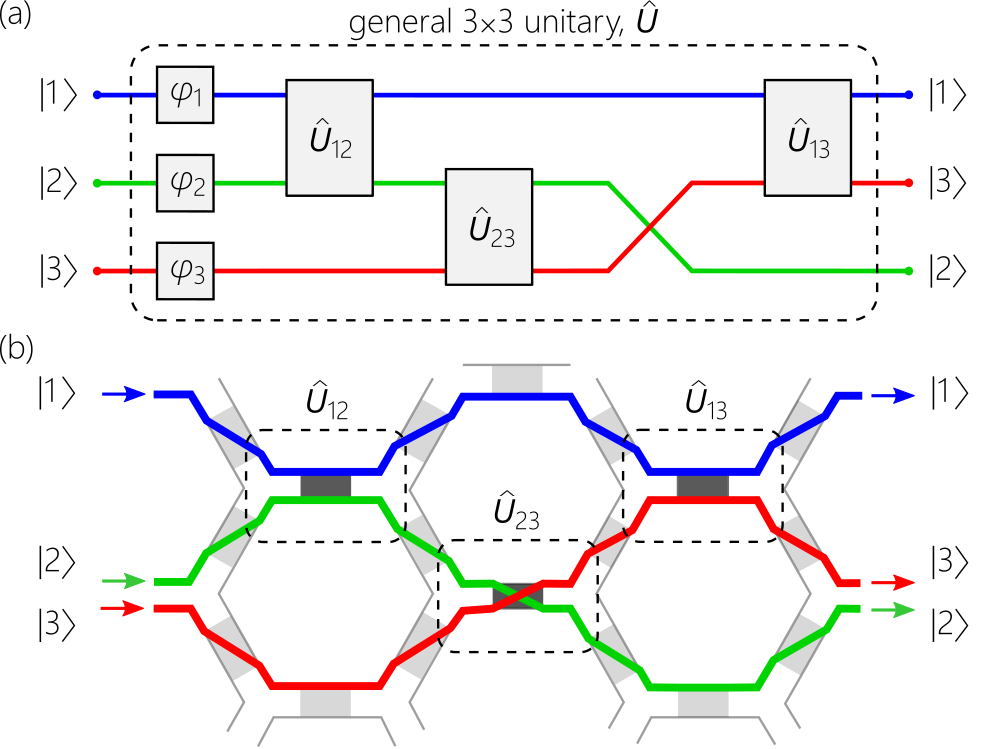}
\caption{(a) A general $3\times3$ unitary is constructed out of a sequence of $2\times2$ unitaries $\hat{U}_{12}$, $\hat{U}_{23}$, and $\hat{U}_{13}$, each operating on a pair of modes, in addition to introducing phases to each of the modes. (b) On-chip layout corresponding to the general $3\times3$ unitary in (a).}
\label{fig:General3x3}
\end{figure}

After tuning the field entropy $S$ by changing the eigenvalues of the input coherence matrix $\mathbf{G}_{\mathrm{o}}=\mathrm{diag}\{\tfrac{1}{3},\tfrac{1}{3},\tfrac{1}{3}\}$ to $\mathbf{G}=\mathrm{diag}\{\lambda_{1},\lambda_{2},\lambda_{3}\}$, we then change its structure via a $3\times3$ unity [Fig.~\ref{fig:General3x3}]. The field entropy $S$ is invariant to the subsequent $3\times3$ unitaries.

The construction of large-dimensional unitaries operating on $N$ modes out of sequences of $2\times2$ unitaries operating on pairs of modes has been studied extensively in quantum information processing \cite{Reck94PRL}. The most general $3\times3$ unitary can be constructed as shown in Fig.~\ref{fig:General3x3}(a): three phase shifters are placed in the modal paths, and then three $2\times2$ unitaries ($\hat{U}_{12}$, $\hat{U}_{23}$, and $\hat{U}_{13}$) are implemented on pairs of modes. The on-chip layout corresponding to the general $3\times3$ unitary is depicted in Fig.~\ref{fig:General3x3}(b).

\section{The modal Stokes-Gell-Mann parameters}

\subsection{Definition of the conventional Stokes parameters}

Consider a partially coherent field supported by two modes $|1\rangle$ and $|2\rangle$ described by the Hermitian, unity-trace, positive semi-definite coherence matrix $\mathbf{G}$:
\begin{equation}\label{eq:2x2G}
\mathbf{G}=\left(\begin{array}{cc}G_{11}&G_{12}\\G_{21}&G_{22}\end{array}\right),
\end{equation}
where $G_{11}=\langle E_{1}E_{1}^{*}\rangle$, $G_{22}=\langle E_{2}E_{2}^{*}\rangle$, and $G_{12}=\langle E_{1}E_{2}^{*}\rangle$; here $E_{1}$ and $E_{2}$ are the field components associated with the modes $|1\rangle$ and $|2\rangle$, respectively, $G_{21}=G_{12}^{*}$, $\mathrm{Tr}\{\mathbf{G}\}=G_{11}+G_{22}=1$. The standard example utilizing this matrix formulation is the polarization degree-of-freedom where $|1\rangle$ and $|2\rangle$ correspond to two orthogonal polarization modes. However, the same description is applicable to a pair of orthogonal spatial modes \cite{Abouraddy19Optica,Halder21OL}.

The conventional SPs $\{s_{0},s_{1},s_{2},s_{3}\}$ are the expansion coefficients of the coherence matrix $\mathbf{G}$ in terms of the Pauli matrices $\{\hat{\sigma}_{0},\hat{\sigma}_{1},\hat{\sigma}_{2},\hat{\sigma}_{3}\}$,
\begin{equation}\label{eq:2x2SPs}
\mathbf{G}=\frac{1}{2}\sum_{j=0}^{3}s_{j}\hat{\sigma}_{j}=\frac{1}{2}\left(\begin{array}{cc}1+s_{1}&s_{2}-is_{3}\\s_{2}+is_{3}&1-s_{1}\end{array}\right),
\end{equation}
where $s_{j}=\mathrm{Tr}\{\hat{\sigma}_{j}\mathbf{G}\}$, $\hat{\sigma}_{0}=\hat{\mathbb{I}}_{2}$, $\hat{\mathbb{I}}_{2}$ is the $2\times2$ identity matrix, $s_{0}=1$, and the Pauli matrices are:
\begin{equation}\label{eq:PauliMatrices}
\hat{\sigma}_{1}=\left(\begin{array}{cc}1&0\\0&-1\end{array}\right),\;
\hat{\sigma}_{2}=\left(\begin{array}{cc}0&1\\1&0\end{array}\right),\;
\hat{\sigma}_{3}=\left(\begin{array}{cc}0&-i\\i&0\end{array}\right).
\end{equation}
The critical properties of the Pauli matrices in the context of modal SPs are $\mathrm{Tr}\{\hat{\sigma}_{j}\}=0$ for $j\neq0$ and $\mathrm{Tr}\{\hat{\sigma}_{j}\hat{\sigma}_{k}\}=2\delta_{jk}$ (excluding $j=k=0$). This allows us to extract the modal SPs from $\mathbf{G}$ via the projection $s_{j}=\mathrm{Tr}\{\hat{\sigma}_{j}\mathbf{G}\}$.

To measure the SPs, the two modes traverse one of three unitaries $\hat{U}_{j}$ ($j=1,2,3$) before the modal weights $I_{1}^{(j)}=G_{11}$ and $I_{2}^{(j)}=G_{22}$ are recorded, where:
\begin{equation}\label{eq:UnitariesU2U3}
\hat{U}_{1}=\hat{\mathbb{I}}_{2},\;\hat{U}_{2}=\frac{1}{\sqrt{2}}\left(\begin{array}{cc}1&1\\-1&1\end{array}\right),\;
\hat{U}_{3}=\frac{1}{\sqrt{2}}\left(\begin{array}{cc}1&-i\\-i&1\end{array}\right).
\end{equation}
After traversing $\hat{U}_{1}=\hat{\mathbb{I}}_{2}$, $s_{1}=I_{1}^{(1)}-I_{2}^{(1)}$; after traversing $\hat{U}_{2}$, $s_{2}=I_{1}^{(2)}-I_{2}^{(2)}$; and after traversing $\hat{U}_{3}$, $s_{3}=I_{1}^{(3)}-I_{2}^{(3)}$. We normalize the modal SPs to $s_{0}=I_{1}^{(j)}+I_{2}^{(j)}=G_{11}+G_{22}$, which is the same in any of these three configurations \cite{Hashemi26arxiv}.

\begin{figure*}[t!]
\centering
\includegraphics[width=18.4cm]{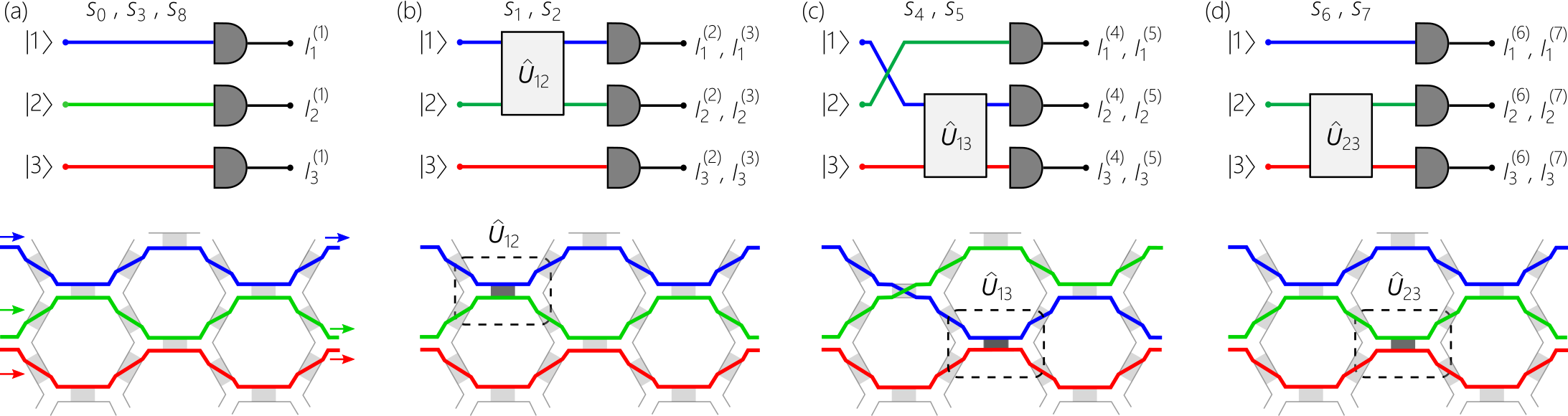}
\caption{Configurations to measure the SGM parameters. Under each conceptual scheme we plot the corresponding layout of the modal paths in the on-chip MZI mesh. The paths followed by the modes are highlighted in different colors. The dashed rectangle in the lower panel identifies the portion of the integrated circuit corresponding to the $2\times2$ unitary depicted in the upper panel. (a) The SGM parameters $s_{0}$, $s_{3}$, and $s_{8}$ are acquired by measuring the modal weights directly (Eq.~\ref{eq:DiagonalSGM}). (b) To acquire $s_{1}$ and $s_{2}$, a unitary $\hat{U}_{12}$ is implemented on modes $|1\rangle$ and $|2\rangle$: $\hat{U}_{12}=\hat{U}_{2}$ to obtain $s_{1}$  and $\hat{U}_{12}=\hat{U}_{3}$ to obtain $s_{2}$. (c) To acquire $s_{4}$ and $s_{5}$, a unitary $\hat{U}_{13}$  is implemented on modes $|1\rangle$ and $|3\rangle$: $\hat{U}_{13}=\hat{U}_{2}$ to obtain $s_{4}$  and $\hat{U}_{13}=\hat{U}_{3}$ to obtain $s_{5}$. (d) To acquire $s_{6}$ and $s_{7}$, a unitary $\hat{U}_{23}$  is implemented on modes $|2\rangle$ and $|3\rangle$: $\hat{U}_{23}=\hat{U}_{2}$ to obtain $s_{6}$  and $\hat{U}_{23}=\hat{U}_{3}$ to obtain $s_{7}$. }
\label{fig:GMSConfiguartions}
\end{figure*}

\subsection{Modal SGM parameters for three-mode light}

The question we tackle here is as follows: how does one generalize the definition of the modal SPs (Eq.~\ref{eq:2x2SPs}), commonly used with a $2\times2$ coherence matrix, to its $3\times3$ counterpart? Such an extension requires first generalizing the $2\times2$ Pauli matrices (Eq.~\ref{eq:PauliMatrices}) to accommodate three modes. This can be done by adopting the 8~Gell-Mann matrices, which are $3\times3$ matrices that are the generators of the group $\mathrm{SU}(3)$, just as the Pauli matrices are the generators of the group $\mathrm{SU}(2)$. In optics, the Gell-Mann matrices have been studied in the context of non-paraxial polarization \cite{ALonso23AOP}, where three field components are relevant rather than two as in the paraxial regime \cite{Brosseau06PO}.  

The Gell-Mann matrices are given explicitly as follows:
\begin{eqnarray}\label{eq:GMmatrices}
\hat{\Lambda}_{1}&=&\left(\begin{array}{ccc}0&1&0\\1&0&0\\0&0&0\end{array}\right),
\hat{\Lambda}_{2}=\left(\begin{array}{ccc}0&-i&0\\i&0&0\\0&0&0\end{array}\right),\nonumber\\
\hat{\Lambda}_{3}&=&\left(\begin{array}{ccc}1&0&0\\0&-1&0\\0&0&0\end{array}\right),
\hat{\Lambda}_{4}=\left(\begin{array}{ccc}0&0&1\\0&0&0\\1&0&0\end{array}\right),\nonumber\\
\hat{\Lambda}_{5}&=&\left(\begin{array}{ccc}0&0&-i\\0&0&0\\i&0&0\end{array}\right),
\hat{\Lambda}_{6}=\left(\begin{array}{ccc}0&0&0\\0&0&1\\0&1&0\end{array}\right),\nonumber\\
\hat{\Lambda}_{7}&=&\left(\begin{array}{ccc}0&0&0\\0&0&-i\\0&i&0\end{array}\right),
\hat{\Lambda}_{8}=\frac{1}{\sqrt{3}}\left(\begin{array}{ccc}1&0&0\\0&1&0\\0&0&-2\end{array}\right),
\end{eqnarray}
and $\hat{\Lambda}_{0}=\hat{\mathbb{I}}_{3}$, with $\mathrm{Tr}\{\hat{\Lambda}_{0}\}=3$. The Gell-Mann matrices have the same critical properties as the Pauli matrices, which makes them useful in defining modal SGM parameters; namely, $\mathrm{Tr}\{\hat{\Lambda}_{j}\}=0$ for $j\neq0$ and $\mathrm{Tr}\{\hat{\Lambda}_{j}\hat{\Lambda}_{k}\}=2\delta_{jk}$, $0\leq j,k\leq8$ (excluding $j=k=0$).

We define the SGM parameters $\{s_{j}\}_{j=0}^{8}$ as the expansion coefficients of the coherence matrix $\mathbf{G}$ in terms of the Gell-Mann matrices:
\begin{equation}
\mathbf{G}=\frac{1}{2}\sum_{j=0}^{8}s_{j}\hat{\Lambda}_{j},
\end{equation}
and the modal SGM parameters are obtained via the projections:
\begin{equation}
s_{j}=\mathrm{Tr}\left\{\hat{\Lambda}_{j}\mathbf{G}\right\}.
\end{equation}
Substituting for the Gell-Mann matrices from Eq.~\ref{eq:GMmatrices}, we obtain the general form of the coherence matrix:
\begin{eqnarray}\label{eq:expansion}
\mathbf{G}=\frac{1}{2}\left(\begin{array}{ccc}
s_{0}+s_{3}+\tfrac{1}{\sqrt{3}}s_{8}&s_{1}-is_{2}&s_{4}-is_{5}\\
s_{1}+is_{2}&s_{0}-s_{3}+\tfrac{1}{\sqrt{3}}s_{8}&s_{6}-is_{7}\\
s_{4}+is_{5}&s_{6}+is_{7}&s_{0}-\tfrac{2}{\sqrt{3}}s_{8}\end{array}\right),
\end{eqnarray}
with $s_{0}=\tfrac{2}{3}$ because $\mathrm{Tr}\{\mathbf{G}\}=1$. The challenge we tackle here is the following: if an unknown $3\times3$ coherence matrix $\mathbf{G}$ corresponding to three-mode partially coherent light is given, how do we measure the SGM parameters and thus reconstruct $\mathbf{G}$?

\section{Measuring the SGM parameters}

We illustrate in Fig.~\ref{fig:GMSConfiguartions} the conceptual schemes corresponding to measurements of the modal SGM parameters and the on-chip layouts for their realization. The \textit{diagonal} Gell-Mann matrices ($\hat{\Lambda}_{0}$, $\hat{\Lambda}_{3}$, and $\hat{\Lambda}_{8}$) and their corresponding modal SGM parameters ($s_{0}$, $s_{3}$, and $s_{8}$) can be extracted from a direct measurement of the modal weights $I_{1}^{(1)}$, $I_{2}^{(1)}$, and $I_{3}^{(1)}$. By guiding the modes to the on-chip detectors directly, we obtain the diagonal elements of $\mathbf{G}$: $I_{1}^{(1)}=G_{11}$, $I_{2}^{(1)}=G_{22}$, and $I_{3}^{(1)}=G_{33}$. From these modal-weight measurements we obtain the following SGM parameters:
\begin{eqnarray}\label{eq:DiagonalSGM}
s_{0}&=&\frac{2}{3}\left\{I_{1}^{(1)}+I_{2}^{(1)}+I_{3}^{(1)}\right\},\nonumber\\
s_{3}&=&I_{1}^{(1)}-I_{2}^{(1)},\nonumber\\
s_{8}&=&\frac{1}{\sqrt{3}}\left\{I_{1}^{(1)}+I_{2}^{(1)}-2I_{3}^{(1)}\right\}.
\end{eqnarray}
In the \textit{second} setting depicted in Fig.~\ref{fig:GMSConfiguartions}(b), a $2\times2$ unitary $\hat{U}_{12}$ is placed in the path of modes $|1\rangle$ and $|2\rangle$. Setting $\hat{U}_{12}=\hat{U}_{2}$ (Eq.~\ref{eq:UnitariesU2U3}) we obtain $s_{1}=I_{1}^{(2)}-I_{2}^{(2)}$, and setting $\hat{U}_{12}=\hat{U}_{3}$ (Eq.~\ref{eq:UnitariesU2U3}) we obtain $s_{2}=I_{1}^{(3)}-I_{2}^{(3)}$.

In the \textit{third} setting depicted in Fig.~\ref{fig:GMSConfiguartions}(c), a $2\times2$ unitary $\hat{U}_{13}$ is placed in the path of modes $|1\rangle$ and $|3\rangle$. Setting $\hat{U}_{13}=\hat{U}_{2}$ we obtain $s_{4}=I_{1}^{(4)}-I_{3}^{(4)}$, and setting $\hat{U}_{13}=\hat{U}_{3}$ we obtain $s_{5}=I_{1}^{(5)}-I_{3}^{(5)}$. In the \textit{fourth} setting depicted in Fig.~\ref{fig:GMSConfiguartions}(d), a $2\times2$ unitary $\hat{U}_{23}$ is placed in the path of modes $|2\rangle$ and $|3\rangle$. Setting $\hat{U}_{23}=\hat{U}_{2}$ we obtain $s_{6}=I_{2}^{(6)}-I_{3}^{(6)}$, and setting $\hat{U}_{23}=\hat{U}_{3}$ we obtain $s_{7}=I_{2}^{(7)}-I_{3}^{(7)}$. Once the modal SGM parameters are acquired, we can reconstruct $\mathbf{G}$ by substitution in Eq.~\ref{eq:expansion}, which corresponds to the third task in Fig.~\ref{fig:OnChipMesh}(h).

\begin{figure*}[t!]
\centering
\includegraphics[width=15cm]{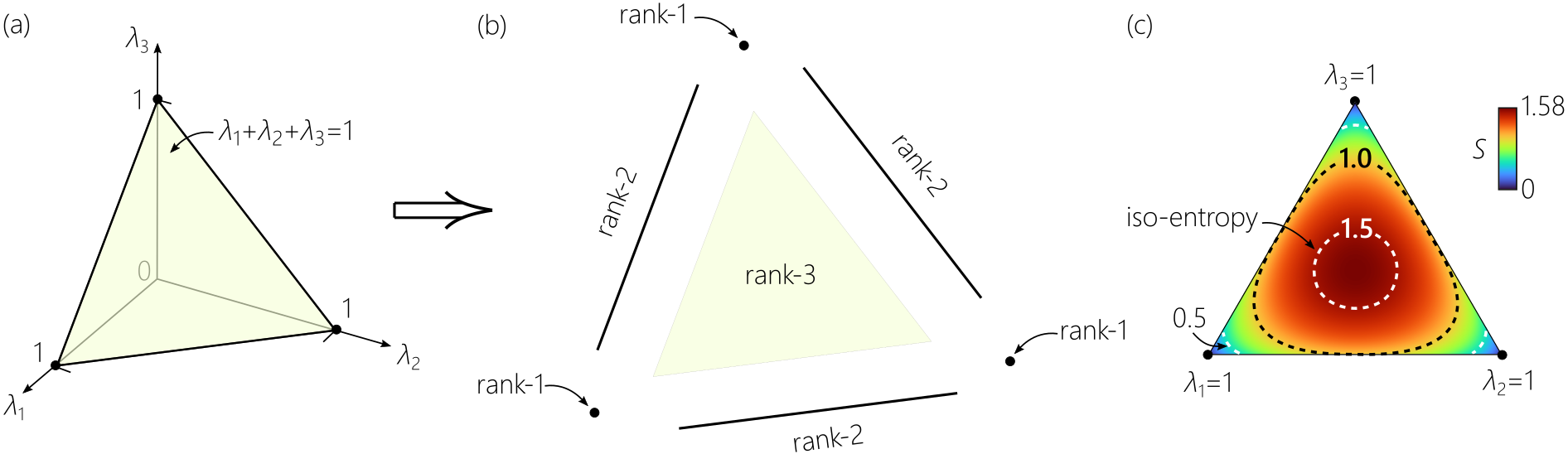} 
\caption{Iso-entropy fields. (a) The surface of the highlighted triangle in $\{\lambda_{1},\lambda_{2},\lambda_{3}\}$-space represents all physically realizable, partially coherent three-mode optical fields described by a $3\times3$ coherence matrix. Each point on the triangle represents a family of iso-entropy coherence matrices that can be inter-converted to each other unitarily. (b) An exploded view of the triangle in (a): the vertices correspond to rank-1 fields, the sides to rank-2 fields, and the area of the triangle to rank-3 fields. (c) Plot of entropy $S$ for $3\times3$ coherence matrices on the surface of the triangle. The contours are iso-entropy curves for selected values of $S$. Different points on each curve represent families of iso-entropy fields that can\textit{not} be inter-converted into each other unitarily.}
\label{fig:EntropyTheory}
\end{figure*}

\section{Measuring the field entropy}

The entropy of the field $S$ is determined solely by the eigenvalues $\lambda_{1}$, $\lambda_{2}$, and $\lambda_{3}$ of $\mathbf{G}$, where $\mathrm{Tr}\{\mathbf{G}\}=\lambda_{1}+\lambda_{2}+\lambda_{3}=1$. This constraint represents a tilted plane in the space spanned by $\{\lambda_{1},\lambda_{2},\lambda_{3}\}$ as shown in Fig.~\ref{fig:EntropyTheory}(a). Because the eigenvalues are limited to the range $0\leq\lambda_{1},\lambda_{2},\lambda_{3}\leq1$, the allowed span on this plane is the triangle depicted in Fig.~\ref{fig:EntropyTheory}(a). Any point on this triangle represents not only a single coherence matrix $\mathbf{G}^{\mathrm{D}}=\mathrm{diag}\{\lambda_{1},\lambda_{2},\lambda_{3}\}$, but also all the other non-diagonal coherence matrices $\mathbf{G}$ that are related to $\mathbf{G}^{\mathrm{D}}$ via a $3\times3$ unitary $\hat{U}$, $\mathbf{G}=\hat{U}\mathbf{G}^{\mathrm{D}}\hat{U}^{\dagger}$, which all share the same value of entropy. In other words, each point in the triangle represents the family of iso-entropy fields that can be inter-converted into each other unitarily.

We have recently shown that the coherence rank (defined as the number of non-zero eigenvalues of the coherence matrix) plays a key role in determining the properties of the field \cite{Harling24PRA,Harling24PRA2,Abouraddy26OL}. For a $3\times3$ coherence matrix, the rank can be~1, 2, or~3. Rank-1 fields once diagonalized have the form $\mathbf{G}=\mathrm{diag}\{1,0,0\}$, which correspond to coherent fields free of random fluctuations ($S=0$). Such fields are represented by the vertices of the triangle in Fig.~\ref{fig:EntropyTheory}(a). Rank-2 fields have the diagonalized form $\mathbf{G}=\mathrm{diag}\{\lambda_{1},\lambda_{2},0\}$ with $\lambda_{1}+\lambda_{2}=1$ and $0<S\leq1$~bit, which correspond to the sides of the triangle in Fig.~\ref{fig:EntropyTheory}(a). Finally, rank-3 fields have the diagonalized form $\mathbf{G}=\mathrm{diag}\{\lambda_{1},\lambda_{2},\lambda_{3}\}$ with $\sum_{j=1}^{3}\lambda_{j}=1$ and $0<S\leq\log_{2}3\approx1.585$~bits. Such fields are represented by the points in the area of the triangle (excluding the vertices and the sides). Thus, the different geometric structures of the triangle in Fig.~\ref{fig:EntropyTheory}(a) correspond to fields of different coherence rank, as highlighted in the exploded view in Fig.~\ref{fig:EntropyTheory}(b).

\begin{figure}[t!]
\centering
\includegraphics[width=8.5cm]{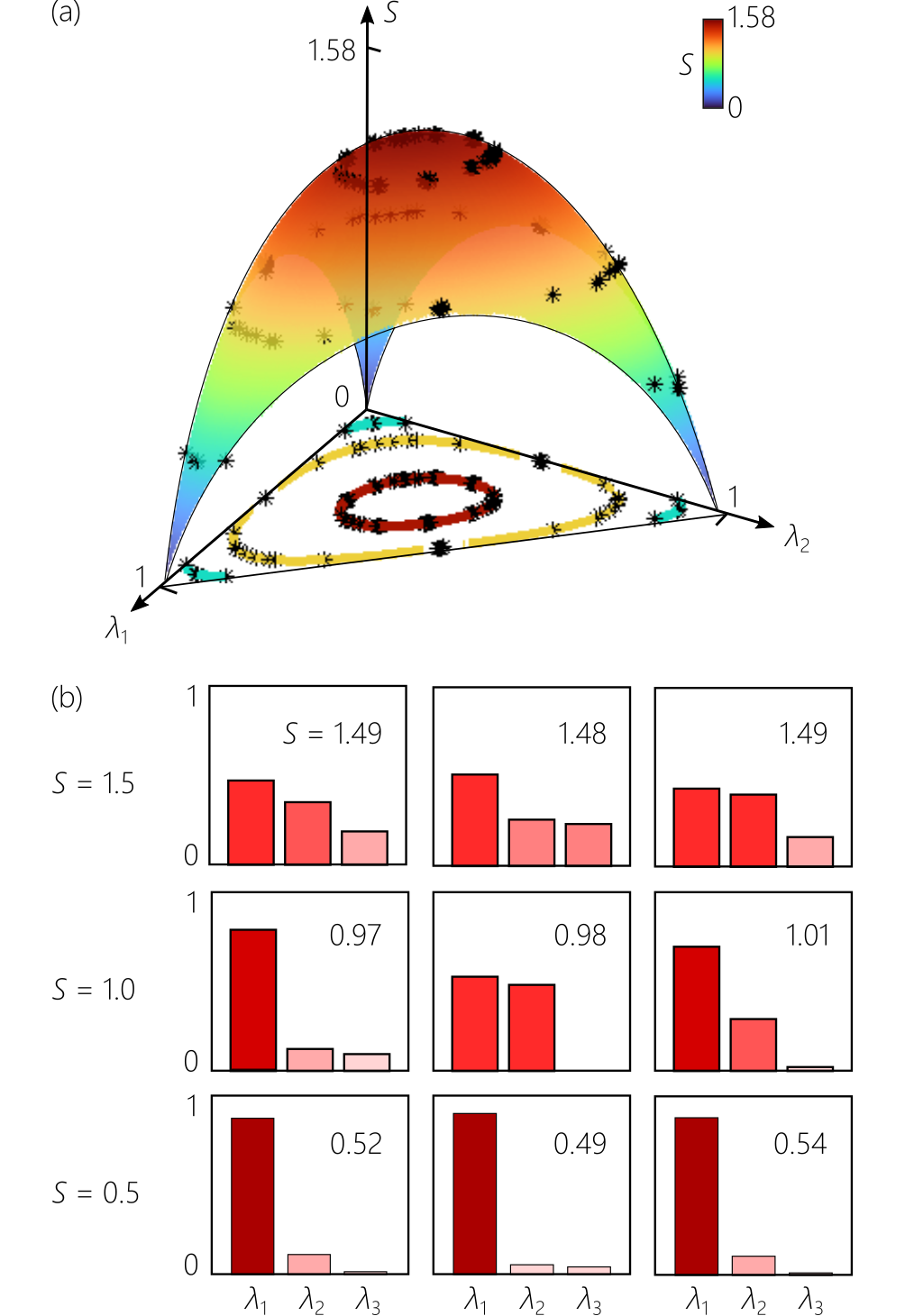}
\caption{(a) Values of $S$ computed from reconstructed coherence matrices after measuring the modal SGM parameters. We plot the data with respect to $\lambda_{1}$ and $\lambda_{2}$ along with the theoretical surface. The coherence matrices correspond to iso-entropy fields with $S=0.5,1.0,$ and 1.5~bits. We project the data onto the ground plane and compare to theoretical iso-entropy curves. (b) Eigenvalues extracted from the reconstructed coherence matrices.}
\label{fig:IsoEntropy}
\end{figure}

As mentioned above, each point on the triangle in the space spanned by $\{\lambda_{1},\lambda_{2},\lambda_{3}\}$ represents the family of all iso-entropy fields that can be inter-converted unitarily (they share the same eigenvalues). Note however that for rank-3 fields where $S=-\sum_{j=1}^{3}\lambda_{j}\log_{2}\lambda_{j}$, the family of iso-entropy fields encompasses those that do not share the same eigenvalues. Indeed, for fixed $S$, along with the unity-trace condition $\sum_{j=1}^{3}\lambda_{j}=1$, we have a one-parameter family of iso-entropy fields that are represented by curves in the triangle, as shown in Fig.~\ref{fig:EntropyTheory}(c). The maximum-entropy field ($S=\log_{2}3$~bits), which is uniquely $\mathbf{G}=\mathrm{diag}\{\tfrac{1}{3},\tfrac{1}{3},\tfrac{1}{3}\}$, corresponds to the point at center of the triangle. Reducing $S<\log_{2}3$~bits corresponds to a curve of increasing circumference. Each point on the curve represents the family of iso-entropy fields that can be inter-converted into each other unitarily, whereas the different points on the curve represent the iso-entropy fields that can\textit{not} be inter-converted into each other unitarily. As long as $S>1$~bit, the iso-entropy curve is entirely enclosed within the triangle and includes only rank-3 fields. When $S=1$~bit, the curve is tangential to the three sides of the triangle. The tangential points correspond to rank-2 fields $\mathbf{G}=\mathrm{diag}\{\tfrac{1}{2},\tfrac{1}{2},0\}$ with $S=1$~bit, while the remainder of the curve corresponds to rank-3 fields. When $S<1$~bit, the iso-entropy curve becomes disjoint and breaks unto three sections in the vicinity of the vertices of the triangle. Whereas the curves still represent iso-entropy rank-3 fields, the termination points on the sides correspond to rank-2 fields that share the same value of entropy. 

\begin{figure*}[t!]
\centering
\includegraphics[width=17.9cm]{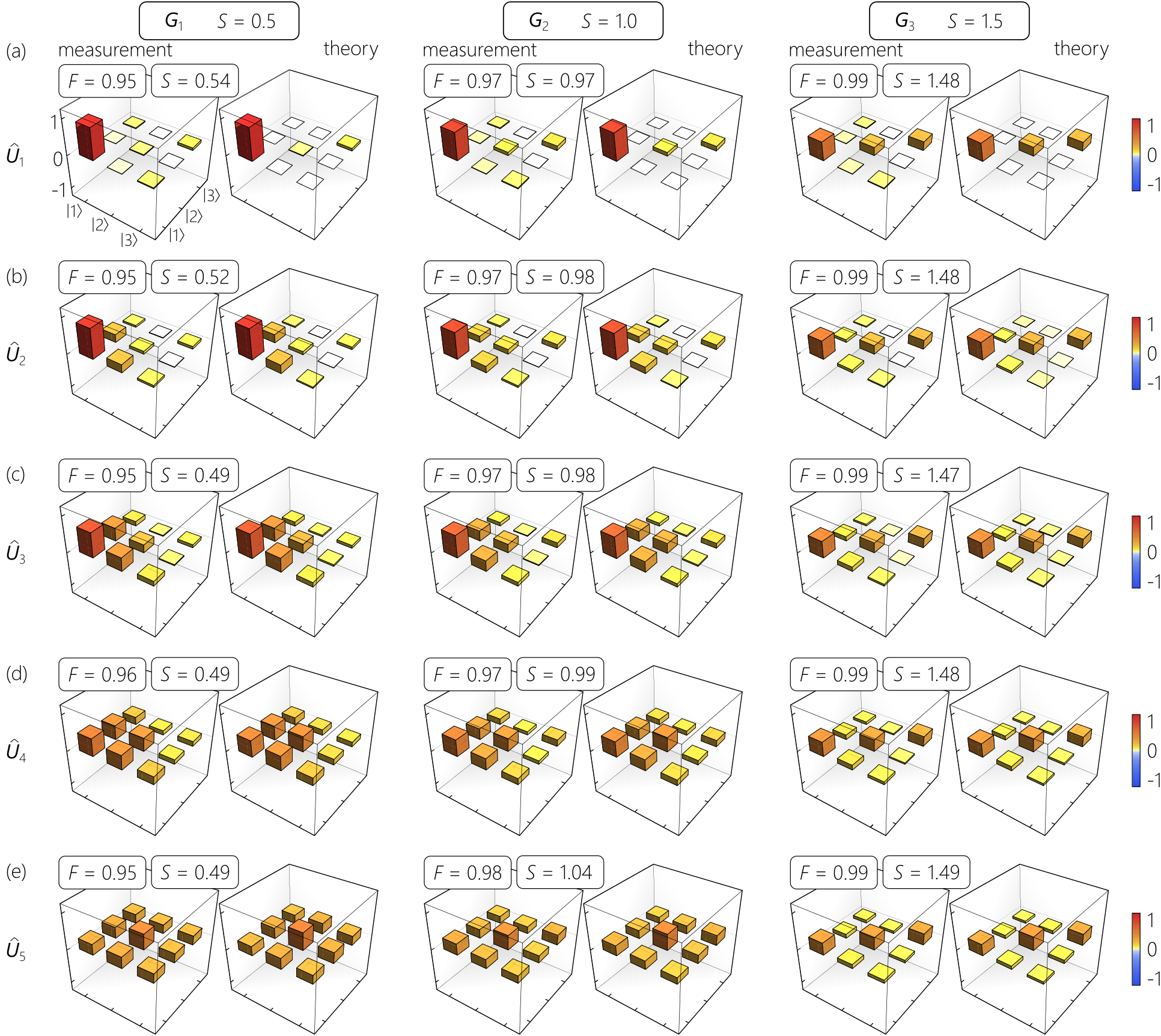} 
\caption{Measurements of iso-entropy fields that are inter-converted into each other unitarily. We plot each measured coherence matrix $\mathbf{G}$ next to the theoretical counterpart. The three double columns correspond to three values of entropy: $\mathbf{G}_{1}$ having $S=0.5$ bit (left); $\mathbf{G}_{2}$ having $S=1.0$~bit (middle); and $\mathbf{G}_{3}$ having $S=1.5$~bits (right). The initial coherence matrices are all diagonal and are given in Eq.~\ref{eq:DiagonalInputs}. Each row corresponds to the implementation of a unitary $\hat{U}_{k}$, $k=1,\cdots,5$, to the coherence matrices with $\hat{U}_{k}=\hat{U}_{12}\hat{U}_{23}(\delta_{k})\hat{U}_{13}$: (a) $\hat{U}_{1}$ with $\delta_{1}=0$; (b) $\hat{U}_{2}$ with $\delta_{2}=\tfrac{\pi}{4}$; (c) $\hat{U}_{3}$ with $\delta_{3}=\tfrac{\pi}{2}$; (d) $\hat{U}_{4}$ with $\delta_{4}=\tfrac{3\pi}{4}$; and (e) $\hat{U}_{5}$ with $\delta_{5}=\pi$. For each reconstructed coherence matrix, we provide the fidelity $\mathcal{F}$ and the estimated entropy $S$.}
\label{fig:UnitaryData}
\end{figure*}

In our measurements, we select the iso-entropy value of $S=1.5,1.0,$ and 0.5~bits. For each value, we scan over the values of $\{\lambda_{1},\lambda_{2},\lambda_{3}\}$ that share the same value of $S$. We detect the modal weights of the resulting coherence matrix $\mathbf{G}=\mathrm{diag}\{\lambda_{1},\lambda_{2},\lambda_{3}\}$ and calculate the associated entropy $S$. In Fig.~\ref{fig:IsoEntropy}(a) we plot the entropy $S$ for all the reconstructed coherence matrices with respect to the values of $\lambda_{1}$ and $\lambda_{2}$, with $\lambda_{3}=1-\lambda_{1}-\lambda_{2}$. We note that all the values of the measured entropy cluster around the iso-entropy curves the selected values of $S$. We plot in Fig.~\ref{fig:IsoEntropy}(b) the eigenvalues computed from selected reconstructed coherence matrices. In general, we find that the estimated value of $S$ is consistently close to the target value. 

\section{Tuning the structure of the coherence matrix}

We now proceed to demonstrate full control over the structure of the coherence matrix $\mathbf{G}$ by tuning its structure via a $3\times3$ unitary $\hat{U}$ constructed along the lines of Fig.~\ref{fig:General3x3}(a) followed by measurement of the SGM parameters and reconstruction of $\mathbf{G}$. Because the unitary $\hat{U}$ leaves the entropy of $\mathbf{G}$ invariant, this amounts to exploring iso-entropy fields that \textit{can} be inter-converted into each other unitarily.

We start with three diagonal coherence matrices:
\begin{eqnarray}\label{eq:DiagonalInputs}
\mathbf{G}_{1}&=&\mathrm{diag}\{0.91,0.17,0.073\},\nonumber\\
\mathbf{G}_{2}&=&\mathrm{diag}\{0.77,0.13,0.1\},\nonumber\\
\mathbf{G}_{3}&=&\mathrm{diag}\{0.5,0.25,0.25\},
\end{eqnarray}
which are selected to correspond to entropy values $S=0.5,1.0,$ and 1.5~bits, respectively. Each diagonal coherence matrix is then modified via one of 5~unitaries, each having the structure $\hat{U}=\hat{U}_{13}\hat{U}_{23}\hat{U}_{12}$, corresponding to the layout in Fig.~\ref{fig:General3x3}. We hold $\hat{U}_{12}$ and $\hat{U}_{13}$ fixed while varying $\hat{U}_{23}$. Specifically, we take $\hat{U}_{12}$ and $\hat{U}_{13}$ as follows:
\begin{equation}
\hat{U}_{12}=\tfrac{1}{\sqrt{2}}\left(\begin{array}{ccc}1&1&0\\1&-1&0\\0&0&\sqrt{2}\end{array}\right), \hat{U}_{13}=\tfrac{1}{\sqrt{2}}\left(\begin{array}{ccc}1&0&1\\0&\sqrt{2}&0\\1&0&-1\end{array}\right),
\end{equation}
each of which rotates a modal pair: modes $|1\rangle$ and $|2\rangle$ via $\hat{U}_{12}$, and modes $|1\rangle$ and $|3\rangle$ via $\hat{U}_{13}$. The intervening unitary $\hat{U}_{23}$ is given by:
\begin{equation}
\hat{U}_{23}=\left(\begin{array}{ccc}
1&0&0\\
0&\sin\tfrac{\delta}{2}&\cos\tfrac{\delta}{2}\\
0&\cos\tfrac{\delta}{2}&-\sin\tfrac{\delta}{2}
\end{array}\right).
\end{equation}
Here the angular parameter $\delta$ determines the coupling between the modes $|2\rangle$ and $|3\rangle$. We select the following values: $\delta=0,\tfrac{\pi}{4},\tfrac{\pi}{2},\tfrac{3\pi}{4}$, and $\pi$. The overall unitary is $\hat{U}(\delta)=\hat{U}_{13}\hat{U}_{23}(\delta)\hat{U}_{12}$, which tunes the structure of the coherence matrices $\mathbf{G}_{j}$, $j=1,2,3$, to $\mathbf{G}_{j}'=\hat{U}(\delta)\mathbf{G}_{j}\hat{U}^{\dagger}(\delta)$, giving rise to a total of 15~distinct coherence matrices. We plot in Fig.~\ref{fig:UnitaryData} the coherence matrices reconstructed from measurements of the SGM parameters along with the theoretical coherence matrices.

We evaluate the quality of the reconstruction using the fidelity, a concept borrowed from quantum information processing \cite{Jozsa94JMO}, defined as $\mathcal{F}=(\mathrm{Tr}\{\sqrt{\mathbf{G}_{\mathrm{m}}}\mathbf{G}_{\mathrm{th}}\sqrt{\mathbf{G}_{\mathrm{m}}}\})^{2}$, where $\mathbf{G}_{\mathrm{th}}$ and $\mathbf{G}_{\mathrm{m}}$ are the theoretically expected and the measured coherence matrices, respectively. The fidelity serves as a benchmark to evaluate the performance of the unitary $\hat{U}$. The lowest fidelity of any of the synthesized coherence matrices is~0.95, but is usually in the vicinity of~0.98. Furthermore, we estimate the eigenvalues of all the reconstructed coherence matrices and calculate the corresponding entropy. In all cases, the estimated entropy from the reconstructed $\mathbf{G}$ is within $5\%$ of the target entropy.

\section{Discussion and Conclusion}

The work presented here represents a critical step towards generalizing Stokes tomography to the on-chip reconstruction of coherence matrices of any dimension, thereby extending the potential reach of structured coherence in application involving optical communications and information processing. All previous on-chip experimental efforts reconstructed coherence matrices with even dimension ($N=2$ in \cite{Hashemi26arxiv} and $N=4$ in \cite{Hashemi26arxiv4Modes}). Here we provided the first reconstruction of \textit{odd}-dimensional coherence matrices. The reconstruction via Stokes tomography for $N=3$ necessitates reliance on the Gell-Mann matrices, yielding the first direct measurement of the SGM parameters in optics. This is a crucial step towards factorizing the measurements needed for reconstructing coherence matrices associated with fields having a large odd dimension.

We have measured the modal SGM parameters for fields confined to on-chip single-mode waveguides or external single-mode fibers, so this does not yet solve the long-standing problem of measuring the SGM parameters for non-paraxial 3D partially polarized light. This requires establishing a transduction from the three polarization modes to three on-chip waveguide modes, which we are currently pursuing.

In conclusion, we have presented the first complete measurement of the modal Stokes-Gell-Mann (SGM) parameters for three-mode partially coherent light represented by a $3\times3$ coherence matrix, which have been studied extensively on the theoretical level but never before implemented experimentally. Measuring the modal SGM parameters enables reconstructing the $3\times3$ coherence matrix. We have implemented the measurements on-chip in an integrated photonic platform comprising a hexagonal mesh of MZIs, with 7~distinct measurement configurations enabling measuring all the modal SGM parameters. This efficient reconstruction procedure paves the way to utilizing multi-moded partially coherent optical fields described by large-dimensional coherence matrices in applications of optical communications and signal processing.

\begin{backmatter}
\bmsection{Acknowledgments}
We thank G.~Li and D.~Hudson for loan of equipment, and Saikat Saha for fiber connectorization.

\bmsection{Funding}
U.S. Office of Naval Research (ONR) N00014-20-1-2789.


\end{backmatter}

\bibliography{diffraction}

\end{document}